\documentclass[epjST,final]{svjour3}
\usepackage{graphicx,amsfonts,amsmath,amssymb,xcolor,subfigure}
\usepackage[english]{babel}
\usepackage[numbers,sort&compress]{natbib}
\usepackage[latin1]{inputenc}

\newcommand{\av}[1]{\left\langle #1 \right\rangle}

\newcommand{\bs}[1]{\boldsymbol #1}

\begin{document}
\title{Motion of Euglena Gracilis: Active Fluctuations and Velocity
  Distribution}

\author{P. Romanczuk \and
  M. Romensky \and D. Scholz \and
  V. Lobaskin
   \and
  L. Schimansky-Geier}
\institute{ P. Romanczuk \at
Department of Ecology and Evolutionary Biology, Princeton University, Princeton, New Jersey 08544, USA; Thaer-Institute, Humboldt Universit{\"a}t zu Berlin, 10099 Berlin \and
  M. Romensky \at
  Department of Mathematics, Uppsala University, Box 480,
  Uppsala 75106, Sweden \and 
  D. Scholz \at
  Conway Institute, University College
  Dublin, Belfield, Dublin 4, Ireland \and 
  V. Lobaskin \at School of Physics, Complex
  and Adaptive Systems Lab, University College Dublin, Belfield,
  Dublin 4, Ireland, \\\email{vladimir.lobaskin@ucd.ie}
  \and
  L. Schimansky-Geier \at Department of Physics, Humboldt Universit\"{a}t zu Berlin,
  Newtonstr. 15, 12489 Berlin, Germany, \\\email{alsg@physik.hu-berlin.de}
}

\maketitle

\begin{abstract} 
  We study the velocity distribution of unicellular swimming algae \emph{Euglena gracilis} using optical
  microscopy and theory. To characterize a peculiar feature of the experimentally observed
  distribution at small velocities we use the concept of active fluctuations, which was recently
  proposed for the description of stochastically self-propelled particles
  [Romanczuk, P. and Schimansky-Geier, L., Phys. Rev. Lett. {\bf 106}, 230601 (2011)].
  In this concept, the fluctuating forces arise due to internal random
  performance of the propulsive motor. The fluctuating forces are directed in
  parallel to the heading direction, in which the propulsion acts. In the theory, we introduce the
  active motion via the depot model [Schweitzer {\em et al.}, Phys. Rev. Lett. {\bf 80}, 23, 5044 (1998)]. We demonstrate that the
  theoretical predictions based on the depot model 
  with active fluctuations are consistent with the experimentally
  observed velocity distributions. In addition to the model with additive active noise, we obtain
  theoretical results for a constant propulsion with multiplicative noise.
\end{abstract}

\section{Introduction}

The development of the theory of Brownian motion, connected to such famous names as Einstein \cite{einstein1905},
Langevin \cite{langevin1908} or Smoluchowski \cite{smoluchowski1906}, was one of the greatest triumphs in
statistical physics at the beginning of the 20$^\text{th}$ century.
Since then Brownian motion not only assumed a central role in
the foundation of thermodynamics and statistical physics but continues
to date to be a major interdisciplinary research topic.

Recently there has been an increasing focus in biology, physics and chemistry, on so-called {\em active matter} systems, both from experimental and theoretical point of view.
Examples of such far-from-equilibrium systems range from the dynamical
behavior of individual units such as Brownian motors
\cite{reimann2002}, motile cells
\cite{friedrich2007,teeffelen2008,selmeczi2008,bodeker2010,li_persistent_2008,dilao_chemotaxis_2013},
macroscopic animals \cite{niwa1994,komin2004,bazazi2010} or artificial
self-propelled particles
\cite{paxton2004,howse2007,ruckner2007,kumar2008,tierno2010} to large
coupled ensembles of such units and their large scale collective
dynamics
\cite{vicsek1995,couzin2002,chate2006,buhl_disorder_2006,sokolov2007,baskaran2009,romanczuk.p:2012,marchetti.c:2013}.
Major advances in {\em active matter} research have been made possible by the
continuously improving experimental techniques such as for example
automated digital tracking \cite{sokolov2007,selmeczi2008, bazazi2010,bodeker2010}
or the realization of active granular and colloidal
systems \cite{paxton2004,howse2007,kudrolli2008,tierno2010}.

Despite these recent advances, there are still many
open questions regarding the universal properties of active
motion. Already at the level of individual active particles it is important to distinguish thermal
fluctuations resulting from the interactions with a heat bath from effective stochastic forces which have their
origin in the active nature of the system. Recently, it was shown that such active fluctuations may lead
to a complex behavior of the mean squared displacement of self-propelled particles with multiple crossovers \cite{peruani2007,romanczuk.p:2012}, and may manifest themselves in characteristic behavior of probability density functions at small speeds, strikingly different from the ones obtained for thermal fluctuations \cite{romanczuk.p:2011,romanczuk.p:2012}. Nontrivial motion statistics and patterns also arise in systems that can switch between two or more locomotion modes like run-and-tumble \cite{Theves2013,Detcheverry2014} or two different speed states \cite{Detcheverry2014,Thiel2012}.

{ We should note that the term ``active fluctuations'' has been used before to refer to non-equilibrium fluctuations of intracellular filament networks (see e.g. \cite{mizuno_nonequilibrium_2007,brangwynne_intracellular_2009,fodor_energetics_2014}), and recently also in the context of flagellar beating \cite{ma_active_2014}. Although all these concepts can be related on a very fundamental level, here we restrict our definition of active fluctuations to the velocity fluctuations of self-propelled agents, which can be described as active Brownian particles \cite{romanczuk.p:2012}.}

In this work we report on experimental observed velocity distribution of
unicellular swimming algae \emph{Euglena gracilis}, which cannot be
explained by models with passive noise only. To describe the
motion of the cells, we develop a theory of active Brownian motion
with active fluctuations and derive the corresponding particle velocity
distributions. We demonstrate that the measured velocities are consistent with
the theoretical predictions, which allows for insight into the motion characteristics and
control patterns.

The remainder of the paper is organized as follows: In Section
\ref{sec:exper} we describe the experimental setup used to measure velocity
distributions of \emph{Euglena gracilis}. Further on, in Section \ref{sec:self-prop}, we introduce the
theory and discuss the comparison between the theory and experiment. In addition, we
discuss also a second theoretical model, to demonstrate that multiplicative noise
can be also considered within the active fluctuations framework. Finally, we conclude in
Section \ref{sec:conclusion}.

\section{\label{sec:exper}Experiment}
\subsection{System}

In this work, we studied the motion of unicellular photosynthetic flagellate \emph{Euglena gracilis} Klebs using optical microscopy.
It is known that \emph{Euglena gracilis} is most sensitive to blue light: the photosynthetic efficiency is maximal for 465 nm light
\cite{deihn.b:1969}.  Euglena is a flagellated swimmer (a pusher-type), propelled by a single flagellum.
The motion is not purely phototactic: \emph{Euglena} cells in dark conditions perform random motion with much spinning
\cite{osaza.k:2014}.
The threshold intensity of blue light required to cause the negative phototaxis is approximately $10$
mW/cm$^2$ \cite{Ntefidou2003}. This is an almost two times lower limiting light intensity than that for a normal daylight \cite{ozasa.k:2013}. In our
experiments we investigated behavior of the cells in strong blue light beyond the photosynthetic limit when the photophobic behavioural
reactions are most pronounced. The cells were supplied by Carolina Biological Supply Company (Burlington, NC, USA) and grown on an aerated,
inorganic salt medium at room temperature under a 12:12 hours light-dark cycle. The average cell size (length) was about 40 $\mu$m.

\begin{figure}
\centering
\includegraphics[width=8.0cm,clip]{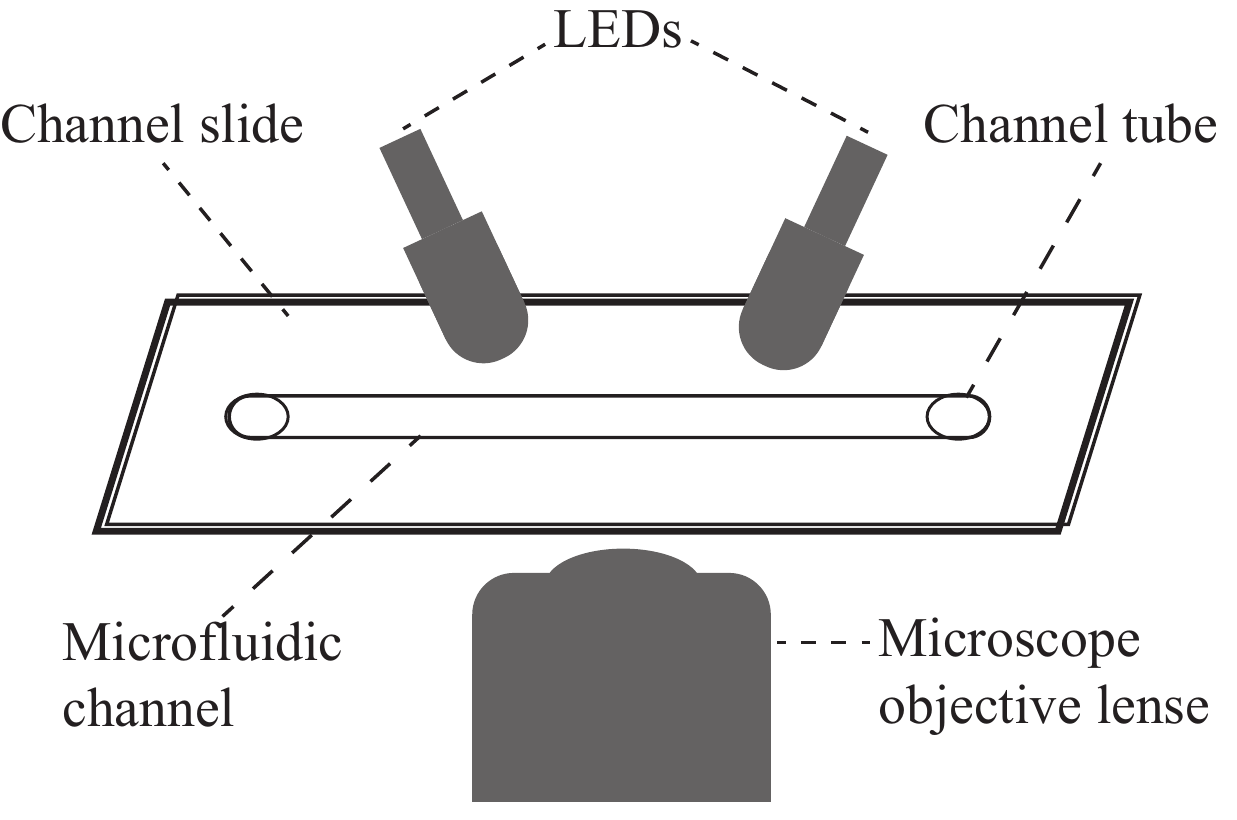}
\caption{Scheme of the experimental setup. %
\label{fig:setup}}
\end{figure}

The schematic of the experimental setup is shown in Fig.~\ref{fig:setup}. During each experimental trial
the cells were confined inside a microfluidic channel ($50$ mm length,
$5$ mm width, $400$ $\mu$m height and $100$ $\mu$l volume) of an
ibiTreat 0.4 Luer channel slide installed into a slide-mount on the
stage of the inverted microscope. Small height of the channel allowed
us to consider motion in two spatial dimensions. The tubes on both
ends of the channel were kept open during the experiment ensuring
proper gaseous cell metabolism. Illumination of the cells with light
was performed by two light emitting diodes (LED) positioned in the
central part of the channel so that the angle between the longitudinal
axis of each LED and the channel plane was fixed at $60^\circ$ and the
distance between the centers of diodes was 10 mm. This arrangement
takes into account characteristics of the directivity pattern of each
LED and ensures uniformity of the illumination within the view field
of the optical system. In this work, we used a pair of blue LEDs with a peak emission
at 465 nm (NSPB500S, Nichia Corporation , Japan). The diodes were lighted up simultaneously. The light intensity
was set to $50$ mW/cm$^{2}$ -- well beyond the photosynthetic
limit. The activation
light from the diodes has been used for detection, no additional
illumination was required. All experiments were performed at room
temperature. Duration of each run was about 10 minutes.
%
%

\subsection{Imaging and data collection}

\begin{figure}[t]
  \centering
  \includegraphics[width=0.5\linewidth]{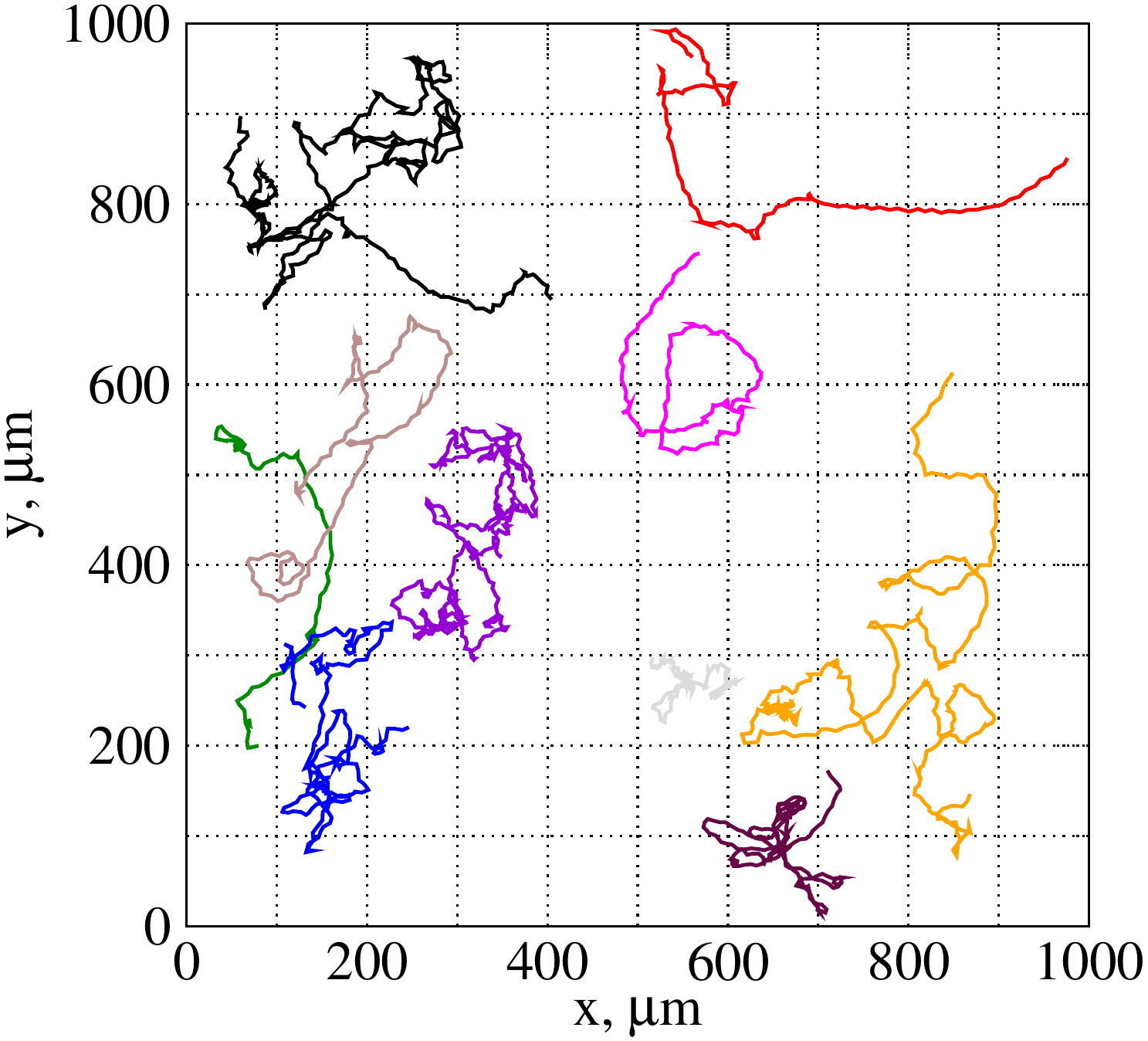}
  \caption{Typical trajectories of {\it Euglena gracilis} cells captured in strong blue light. Only ten trajectories are shown for the sake of clarity.}
  \label{fig:traject}
\end{figure}

The optical readout system consisted of an image acquisition equipment
and data processing computer. Real time images of \emph{Euglena} cells
in a channel slide were captured by the Andor iXon3 885 EMCCD camera ($1004 \times 1004$ pixels, pixel size
8 $\mu$m, 2 fps frame rate) mounted on the optical microscope Zeiss Axiovert 200M
through a 5x objective lens.
 The observation field was 1.6 mm $\times$ 1.6 mm.
 The number of cells within
the observation area was $800-1200$ giving packing fractions of
$0.026-0.039$. Images were recorded in miltipage TIFF format using
original camera manufacturer software. Typical trajectories of the cells are shown in Fig. \ref{fig:traject}.

Further image processing involving cell recognition and tracking was
performed with Imaris package, v7.6 (Bitplane, UK). The raw data
consisted of $x$ and $y$ coordinates of the cell geometrical centers, identities of cells and
a time stamp. These data did not include positions of non-motile (dead) cells which were excluded
with a filter based on a minimal trajectory length (50 $\mu\texttt{m}$).
In the videos, the frame rate seems to be sufficiently high. The displacements between the frames
are less than the cell size. The positional statistics
was subsequently exported into ASCII format files and analyzed with our code.
Velocities of the cells were calculated based on the two subsequent positions. The motion
statistics were averaged over two runs. Due to isotropy of the motion, we
combined the histograms in $x$ and $y$ directions. To extract the slice of
the histogram corresponding to one velocity component $(v_x,0)$, we used
the data where the corresponding transverse component $v_y$ was small
$-0.5\ \mu\text{m/s} < v_y < 0.5\ \mu\text{m/s}$.

\begin{figure}[t]
  \centering
  \includegraphics[width=0.5\linewidth]{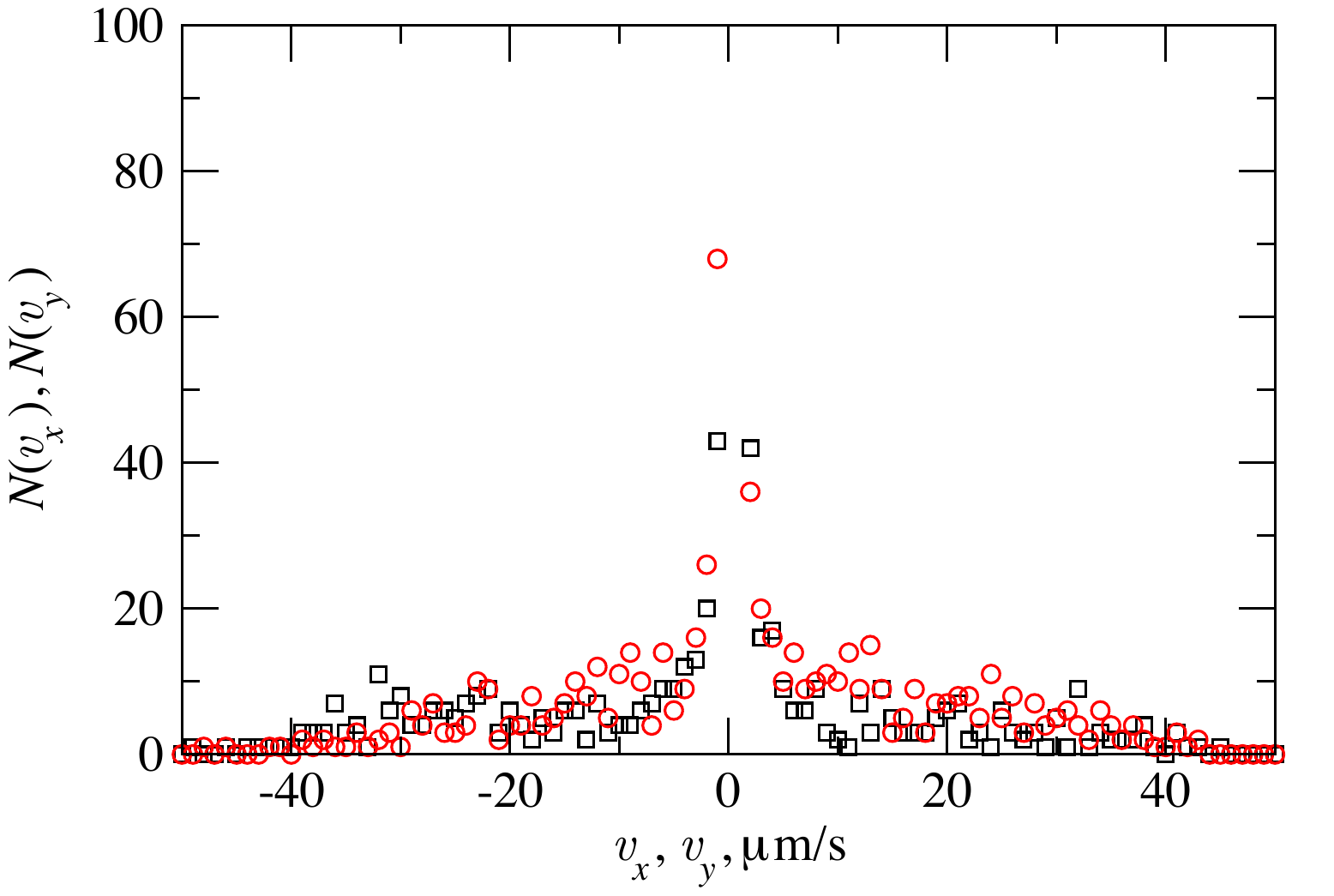}\includegraphics[width=0.5\linewidth]{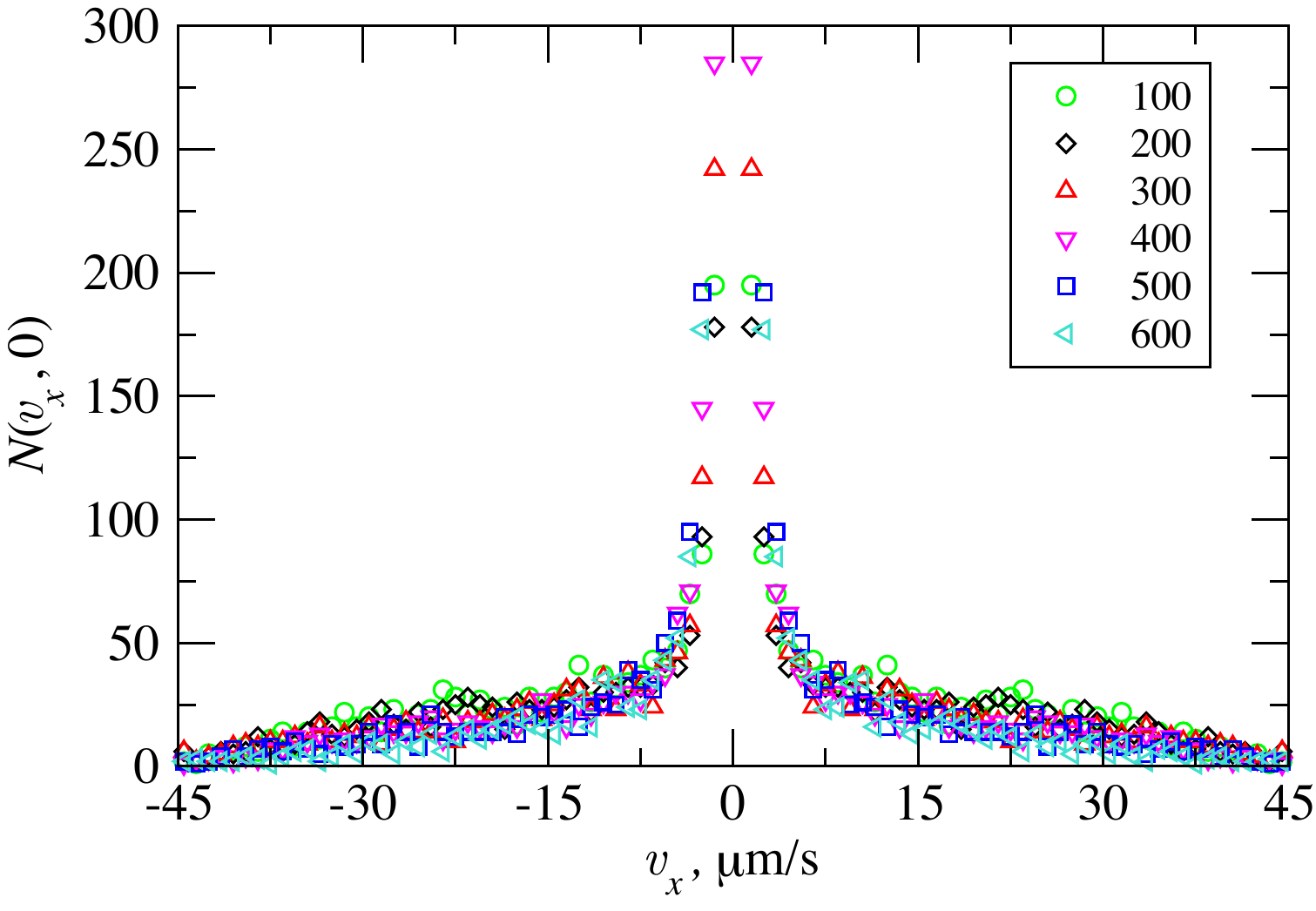}
  \caption{(a) An example of raw data set for {\it Euglena gracilis} in strong blue light collected 200-300 seconds from the start of the illumination.
    (b) Six symmetrised datasets, where $\pm v_x$ and $\pm v_y$ are combined.}
  \label{fig:stats}
\end{figure}

The cells in strong blue light were in constant motion, although
slowing down from time to time. As we noticed the change of the
motion within the duration of the experiment, we split the statistics
into 6 subsets corresponding to time intervals of 0--100, 100--200,
200--300, 300--400, 400--500, and 500--600 seconds.  An example of a
raw dataset with $v_x$ and $v_y$ velocity components for one interval
and the collection of symmetrised data are shown in
Fig. \ref{fig:stats}. We can observe straight away a striking feature
of these distributions: A strong peak around zero
velocity. This is in contrast with the shape of
  distribution predicted for active Brownian particles effected by
  thermal agitation, likewise usual Brownian motion which have a
  crater like shape with a minimum around zero. This agitation from
  the molecules of the surrounding liquid is negligible compared to
  the phototactic propulsion. To understand the peak at vanishing
  velocity we need to introduce a different noise model, which we
present in the next section.

\section{\label{sec:self-prop}Self propelled particles with active fluctuations}
\subsection{\label{sec:active} Models with active fluctuations}
In recent publications two of us have investigated the difference
between passive and active noise acting on the motion of self
propelled units \cite{romanczuk.p:2011,romanczuk.p:2012}. Passive
fluctuations have their origin in the fluctuating environment in which
the particle moves. In result, the effect of a passive random force
$\bs{\eta}_\text{p}(t)$ is independent on the direction of motion
(heading) of the particle. The classical example of passive
fluctuations is ordinary Brownian motion in thermal equilibrium, where
the stochastic force is associated with random collisions of the
Brownian particle with molecules of the surrounding fluid.

In contrast, active fluctuations $\bs{\eta}_\text{a}(t)$ are a pure
far from equilibrium phenomenon. They may become important in the motion of an
biological agents or self-propelled particles. The origin of these
fluctuations can be for example variations in the propulsion of
chemically powered colloids \cite{paxton2004,howse2007,ruckner2007},
complex intra-cellular processes in cell motility
\cite{selmeczi2008,bodeker2010} or unresolved internal decision
processes in animals \cite{niwa1994,komin2004,bazazi2010}.

In this manuscript we show that the concept of active fluctuations is
able to explain measurements of the velocity distribution function of
the \emph{Euglena} cells. Especially, an observed giant peak in the
two dimensional velocity distribution of the Cartesian velocity
components $P(v_x,v_y)$ at the origin $v_y=0,v_y=0$ is due to the
special selection of the noise sources. Only active fluctuations describe
correctly the large peak
in two-dimensional Cartesian velocity distributions at small speeds.

We assume a two-dimensional motion of the cells is given by the
position vector $\bs{r}(t)$ and the velocity $\bs{v}(t)$. Components
are labeled as ${\bf r}=(x,y)$ for the position vector of the cell,
and ${\bf v}=(v_x,v_y)$ being its velocity vector. The dynamics of the
kinematics of the cell is given by the second Newton law. We assume a
propulsive mechanism with force $\bs{a}_p$, Stokes friction and a
random forces $\bs{\eta}(t)$ acting on the cell. In particular, the
dynamics reads afterwards.
\begin{align}
  \label{eq:gv}
  \dot {\bf r} & = {\bf v} \\
  m\dot {\bf v} & = \bs{a}_p -\gamma {\bf v}+\bs{\eta}(t)
\end{align}
and we will set the mass to unity $m=1$, later on.

Active particles have a preferred direction of motion (heading)
determined by their propulsion mechanism (``head-tail'' asymmetry).
The heading vector is a unit vector ${\bf e}_h$ with $|{\bf e}_h|=1$
defining the orientation of the particle along which the propulsive
force acts. This force is independently on other acting external
forces. It acts generally in direction of the instantaneous
velocity, and can be decelerating or accelerating.

\begin{figure}[t]
  \centering
  \includegraphics[width=0.5\linewidth]{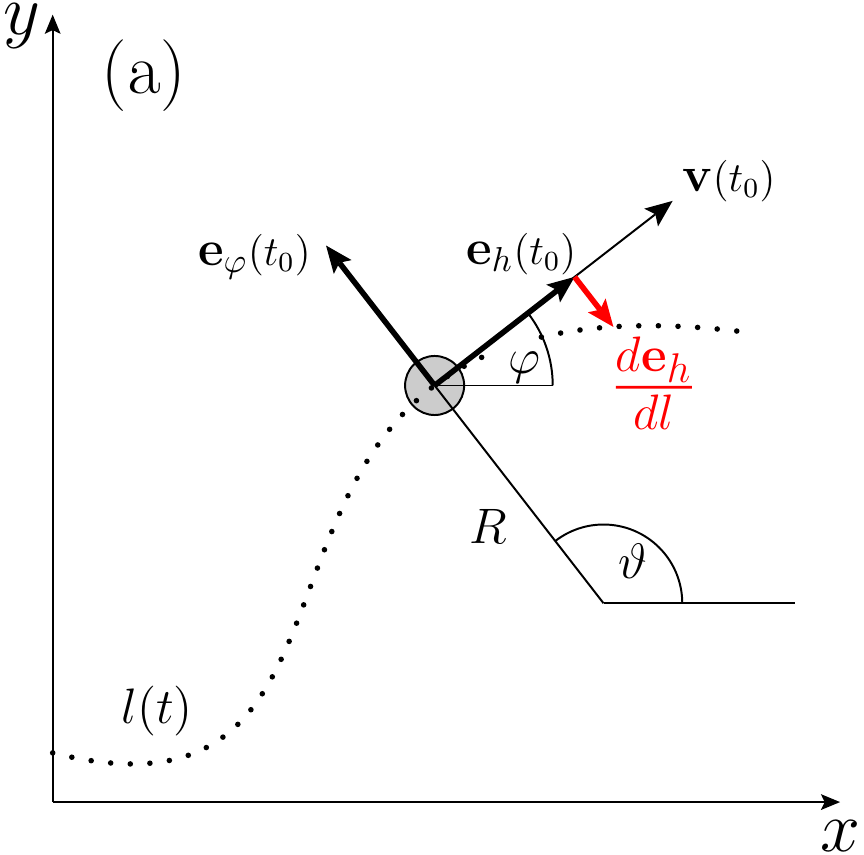}\includegraphics[width=0.25\linewidth]{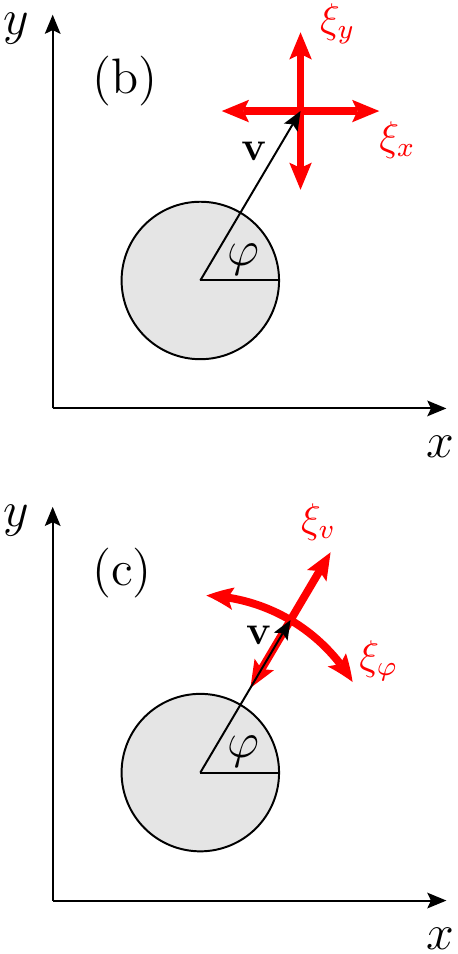}
  \caption{(a) Schematic visualization of motion in a plane with unit
    vectors $\bs{e}_h(t), \bs{e}_{\varphi}(t)$ (thick arrows),
    velocity vector $\bs{v}(t)$ (thin arrow) and ${\rm d}\bs{e}_h/{\rm d}l$
    (thick red/gray arrow). The trajectory of the particle $l(t)$ is
    indicated by the dotted line.  (b) Visualization of passive
    fluctuations $\xi_x(t)$ and $\xi_y(t)$ (thick red/gray
    arrows). (c) Visualization of active fluctuation $\xi_v(t)$ and
    $\xi_{\varphi}(t)$ (thick red/gray arrows). Figure reproduced from \cite{romanczuk_active_2011}.}
  \label{fig:scheme_2dmotion}
\end{figure}

In two spatial dimensions ($d=2$) it is fully determined by the angle
$\varphi$ defining the direction with respect to the $x$-axis (see
Fig. \ref{fig:scheme_2dmotion}a): ${\bf
  e}_h(t)=(\cos\varphi(t),\sin\varphi(t))$, and the evolution of the
position of the particle can be rewritten as
\begin{align}
  \dot {\bf r}(t) & = {\bf v}(t) = v(t) {\bf e}_h(t).
\end{align}
Note, that $v(t)$, which is the projection of the velocity onto the
heading direction, can also adopt negative values if moving backwards
with respect to the propulsive mechanism. The heading vector differs
from the unit vector of the velocity ${\bf e}_v(t)$ to which it is
either parallel or anti-parallel if moving forward or backward.
Projected on ${\bf e}_v(t)$, the velocity is the speed and always
positive and the length of the vector ${\bf v}(t) = s(t) {\bf e}_v(t)$
with $s(t)=\sqrt{v_x^2+v_y^2}=|\bs{v}(t)|$.

The temporal evolution of the velocity vector in the new coordinates
($v$,$\varphi$) reads
\begin{align}
  \dot {\bf v}&=\dot v {\bf e}_h + v \dot \varphi {\bf e}_\varphi \, ,
\end{align}
where ${\bf e}_\varphi=(-\sin\varphi(t),\cos\varphi(t))$ is the unit
vector in the angular direction.  Multiplying Eq. (2) with
${\bf e}_v$ and ${\bf e}_\varphi$, respectively, yields:
\begin{align}
  \label{eq:headingdyn}
  \dot{v}&= a_p -\gamma v + \bs{\eta} \cdot\bs{e}_h \, , \\
  \dot{\varphi} &= \frac{1}{v} \bs{\eta} \cdot \bs{e}_{\varphi} \, .
\end{align}
Here, $\cdot$ indicates a scalar product. One has to note that the
orientational dynamics diverge for $v = 0$. This is due to the fact
that for a point-like particle, as considered here, the polar angle $\varphi$ cannot be defined for a vanishing
speed.

At this point one has to distinguish between the action of active vs
passive noise. Passive noise as in case of Brownian motion acts
independently in the $x$ and $y$ directions $\bs{e}_x$, respectively
$\bs{e}_y$, as Gaussian white noise. It decomposes into two
independent components with
\begin{align}
\label{eq:passivenoise}
\bs{\eta}_\text{p}(t)= \sqrt{2 D} ( \xi_x(t) \bs{e}_x +\xi_y(t)\bs{e}_y),
\end{align}
with following properties
\begin{align}
\label{eq:noise}
\av{\xi_i(t)}=0,\,\,\,\,\av{\xi_i(t)\xi_j(t^\prime)}=\delta_{i,j}\delta(t-t^\prime)
\,\,\,,\quad i,j\in \{x,y\}.
\end{align}
Thus with this choice of velocity variables, the problem becomes a
model with multiplicative noise in \eqref{eq:headingdyn}. It is easier
solved in a Cartesian representation of the velocity vectors, where for passive noise the motion in $x$ and $y$ becomes independent.

In contrast to passive noise, for active fluctuations we assume Gaussian noise
acting with intensity $D_v$ along the heading
direction. It is complemented by a second Gaussian noise randomly changing
the heading direction with intensity $D_{\varphi}$, by acting only
perpendicularly to the heading vector. Therefore, we define
\begin{align}
  \label{eq:activenoise}
  \bs{\eta}_\text{a}(t)= \sqrt{2 D_v} \xi_v(t)\bs{e}_h(t) + \sqrt{2
    D_{\varphi}} \xi_{\varphi}(t)\bs{e}_{\varphi}(t).
\end{align}
Again both noise sources shall be independent
\begin{align}
\label{eq:noise_a}
\av{\xi_i(t)}=0,\,\,\,\,\av{\xi_i(t)\xi_j(t^\prime)}=\delta_{i,j}\delta(t-t^\prime)
\,\,\,, \quad i,j\in \{v,\varphi\}.
\end{align}
A similar decomposition happens in case of active noise compared to
the Cartesian formulation with passive noise. The dynamics along the
heading direction with velocity component $v$ decouple from the
angular motion. This stochastic
$\varphi$-dynamics with homogeneous additive noise along circles of
constant $v$ results even in an equipartition along $\varphi$.
Subsequently the $v$ dynamics can be solved. As a result, models with
active fluctuations becomes a new class of analytically tractable
systems.

\subsection{\label{sec:depot} Energy depot}
To explain measurements of the velocity distribution function of the
\emph{Euglena} cells, we will use the depot model as introduced by
Ebeling, Schweitzer and Tilch
\cite{schweitzer.f:1998,ebeling.w:1999,schweitzer.f:2007,romanczuk.p:2012}.
This model traces the propulsion back to a food uptake of the cells which
rate $q$ from the environment. It fills up an internal energy depot
$e_{in}(t)$ of the cell. The internal energy decay with
rate $c$ and can be transformed into kinetic energy with rate $d$, whereby the corresponding term
is proportional to the internal and kinetic energy of the cell. Thus $e_{in}(t)$ obeys
\begin{align}
\label{eq:depot1}
\dot{e}_{in}=q-c e_{in} -d e_{in} v^2 \ .
\end{align}

The gained kinetic energy is transformed into momentum and creates a
propulsion in the Newtonian force balance along the heading direction
as
\begin{align}
\label{depot2}
\dot{v}&=   d e(t) v(t)-\gamma v + \sqrt{2 D_v} \xi_v(t).
\end{align}
{We assume that the angular dynamics are purely stochastic, driven by Gaussian white noise:}
\begin{align}
\label{eq:depot3}
\dot{\varphi}&=\frac{1}{v}\sqrt{2 D_{\varphi}} \xi_{\varphi}(t).
\end{align}
{This assumption implies also that the angular dynamics are not affected by the depot. In general, this might be not the case, for example angular fluctuation could decrease with decreasing energy depot. However, this would only affect the persistence of the cell trajectory and would make no difference to stationary velocity distributions we focus on here.}
The active noise sources $\xi_k(t)$ are defined in Eq. \ref{eq:noise}.

Assuming a {fast relaxation rate of the depot variable,
  the internal energy responds instantaneously to its supply and
  demand. Thus it assumes a stationary value and can be adiabatically eliminated, despite ongoing exchange with the environment} (for a further
discussion including stochastic supply $q$, see
\cite{romanczuk.p:2012}). It yields
\begin{align}
\label{eq:depot4}
e(t)&=\frac{q}{c+dv^2(t)},
\end{align}
and in consequence the heading velocity changes as
\begin{align}
\dot{v}&=  \frac{d q}{c+d v^2}  v-\gamma v + \sqrt{2 D_v} \xi_v(t)\,.
\label{eq:depot5}
\end{align}
Hence, the propulsive acceleration reads in this approximation
\begin{align}
\label{eq:propul_eu}
a_p&=\frac{d q}{c+d v^2} v.
\end{align}
Finally, the depot model becomes
\begin{align}
\dot{v}&= a_p -\gamma v + \sqrt{2D_v} \xi_v(t), ~~~~~~~\dot{\varphi}= \frac{1}{v} \sqrt{2D_{\varphi}} \xi_{\varphi}(t).
\end{align}
and the corresponding Fokker-Planck equation
\begin{align}
\label{eq:fpe_depot}
  \frac{\partial }{\partial t}p_a(v,\varphi,t|v_0,\varphi_0,t_0) &=
  -\frac{\partial}{\partial v}(a_p-\gamma v)p_a + D_v
  \frac{\partial^2}{\partial v^2}p_a+ \frac{D_{\varphi}}{v^2}\frac{\partial^2}{\partial \varphi^2}p_a\,.
\end{align}
with $a_p$ from Eq. \eqref{eq:propul_eu}.

{The propulsion dynamics can be reformulated by
  introducing the stationary propulsion speed in the absence of noise}
  $v_0$ defined as
\begin{align}
\label{eq:depot_v_0}
v_0 &=\sqrt{ \frac{q}{\gamma} - \frac{c}{d}}.
\end{align}
Eq. (\ref{eq:depot5}) becomes then
\begin{align}
  \label{eq:depot6}
  \dot{v}&= \frac{\gamma d}{c+d v^2} v(v_0^2 -v^2 ) + \sqrt{2D_v} \xi_v(t).
\end{align}
In case that parameters obey $\frac{q}{\gamma} - \frac{c}{d} \ge 0$,
$v_0$ exists and is a stable fixed point of the deterministic speed dynamics. Its positive
root $v_0 \ge 0$ is the mean propulsion speed in the small noise
limit. The depot pumps sufficient energy to excite a stationary
motion. In the opposite case, the single stationary stable solution
of the deterministic part of Eq. (\ref{eq:depot5}) is the rest
state ($v_0=0$). The depot is insufficient to generate persistent motion at the
given damping $\gamma$. Thus, in this case particles are driven by the noise alone. Note that
this noise has to be still associated with active fluctuations produced by the propulsion mechanism of the algae as it acts 
only along the heading direction, which is fundamentally different from the case of thermal agitation.

Later on, we estimate the value of $v_0$ from Eq. (\ref{eq:depot_v_0})
using experimentally fitted parameters.

\subsection{\label{sec:additive} Additive active noise}
The asymptotic stationary distribution $\lim_{t_0 \to -\infty}
p_a(v,\varphi,t|v_0,\varphi_0,t_0) = p_a(v)$ becomes independent of the
initial conditions. Also the noisy $v(t)$ dynamics
become independent from the direction $\varphi$ and in the stationary
limit the probability function density decomposes
$p_a(v,\varphi)=p_a(v) p_d(\varphi)$.  The angular distribution
becomes homogeneous in all directions $\varphi$ with
$p_d(\varphi)=1/2\pi$. The remaining probability density function for
the velocity follows from the stationary Fokker-Planck equation
\begin{align}
\label{fpe_stat10}
\frac{\partial}{\partial t} p_a(v) &= 0 = -\frac{\partial}{\partial
  v}(a_p-\gamma v)p_a + D_v \frac{\partial^2}{\partial v^2} p_a
\end{align}
with $a_p$ from Eq.(\ref{eq:propul_eu}).  The latter ordinary
differential equation can be integrated with vanishing probability
flux. Thus, the stationary probability density function of
velocities projected on the heading dynamics reads
\footnote{Probability density functions are defined up to the
  normalization constant.}
\begin{align}
\label{eq:depot_head}
p_a(v)\propto \left(\frac{c}{d}+v^2\right)^{\frac{q}{2D_v}} \exp\left(
-\frac{1}{2D_v}\gamma v^2\right).
\end{align}
From this, by summing the positive and negative projected velocities, we
obtain the speed distribution, i.e. $\tilde{p}(s)=p_a(s=v)+p_a(s=-v)$
\footnote{Note, it follows from this definition that the derivative of
  the speed distribution $\tilde{p}(s)$ vanishes in the limit $s \to
  0$ if $p_a$ is differentiable at $v=0$.}. It reads
\begin{align}
\label{eq:depot_speed}
\tilde{p}(s)=p_a(s=v)+p_a(s=-v) \propto
\left(\frac{c}{d}+s^2\right)^{\frac{q}{2D_v}} \exp\left( -\frac{1}{2D_v}\gamma s^2\right)\,.
\end{align}
Transformation into the Cartesian velocities requires the consideration
of the Jacobi-determinant for the two-dimensional variable
transformation $s,\varphi \to v_x,v_y$. Therefore a giant peak occurs
at the origin reflecting the non-vanishing speed probability density for the ``resting'' situation in the presence of active noise. Finally, we obtain
\begin{align}
\label{eq:depot_cart}
P(v_x,v_y)\propto \frac{1}{\sqrt{v_x^2+v_y^2}}\left(\frac{c}{d}+v_x^2+v_y^2\right)^{\frac{q}{2D_v}} \exp\left( -\frac{1}{2D_v}\gamma (v_x^2+v_y^2)\right)\,.
\end{align}
\begin{figure}[t]
\centering
\includegraphics[width=0.9\linewidth]{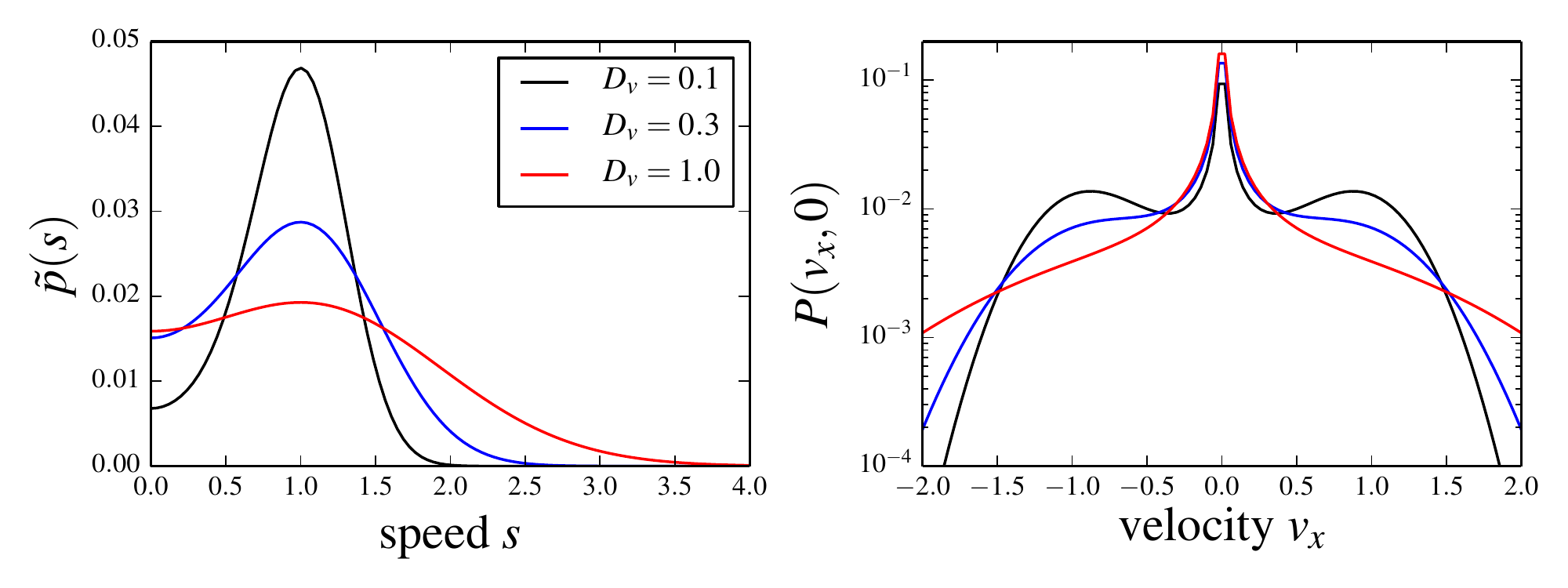}
\caption{The stationary speed probability (Eq.(\ref{eq:depot_speed}), left panel) and
velocity distribution function (Eq.(\ref{eq:depot_cart}),right panel) for the energy depot with additive noise.
Parameters:$q=2 ,d=1, c=1 ,\gamma=1$ and noise intensity $D_v$ as assigned in the inset.
\label{fig:depot}}
\end{figure}

The divergence of the Cartesian probability density is a universal
property of active noise acting along the heading direction \cite{romanczuk.p:2011,romanczuk.p:2012}. It is
intrinsically connected to finite values of the speed probability
density at the origin $\tilde{p}(s=0)>0$. This in turn, is a
consequence of active noise stochastically exploring speed states
infinitely close to the resting situation. These states are never
reached in case of an ordinary Brownian particle, where the permanent
stochastic impacts, independent in $x$- and $y$-direction, always
drive the speed dynamics away from the state $s=0$.

The solution Eq. (\ref{eq:depot_head}) looks very similar to the
stationary solution in the depot model as reported in \cite{erdmann_brownian_2000}. However it has to be emphasized that it describes
a different physical situation. The solution in \cite{erdmann_brownian_2000}
describes the Cartesian velocity components as the results from a
passive noise as given by Eq.(\ref{eq:passivenoise}). In the case
discussed here the solution was obtained along the co-moving heading
direction and Cartesian velocity components are due to Eq.
(\ref{eq:depot_cart}).

As shown in a previous paper \cite{grosmann_active_2012} the
existence of the state with a maximum in the probability density
function can act as a strong attractor in a situation with many
particles and a velocity-alignment interaction between particles. In
this situation the particles with active noise are collectively
``trapped'' in the resting state in contrast if passive noise is
applied. The character of the phase transition between the ordered
(running particles) state and disordered state changes from a second
order to a first order transition due to the coexistence of the
running and resting states.

\subsection{\label{sec:comp} Comparison with measurements}

\begin{figure}
\centering
\includegraphics[width=8.0cm,clip]{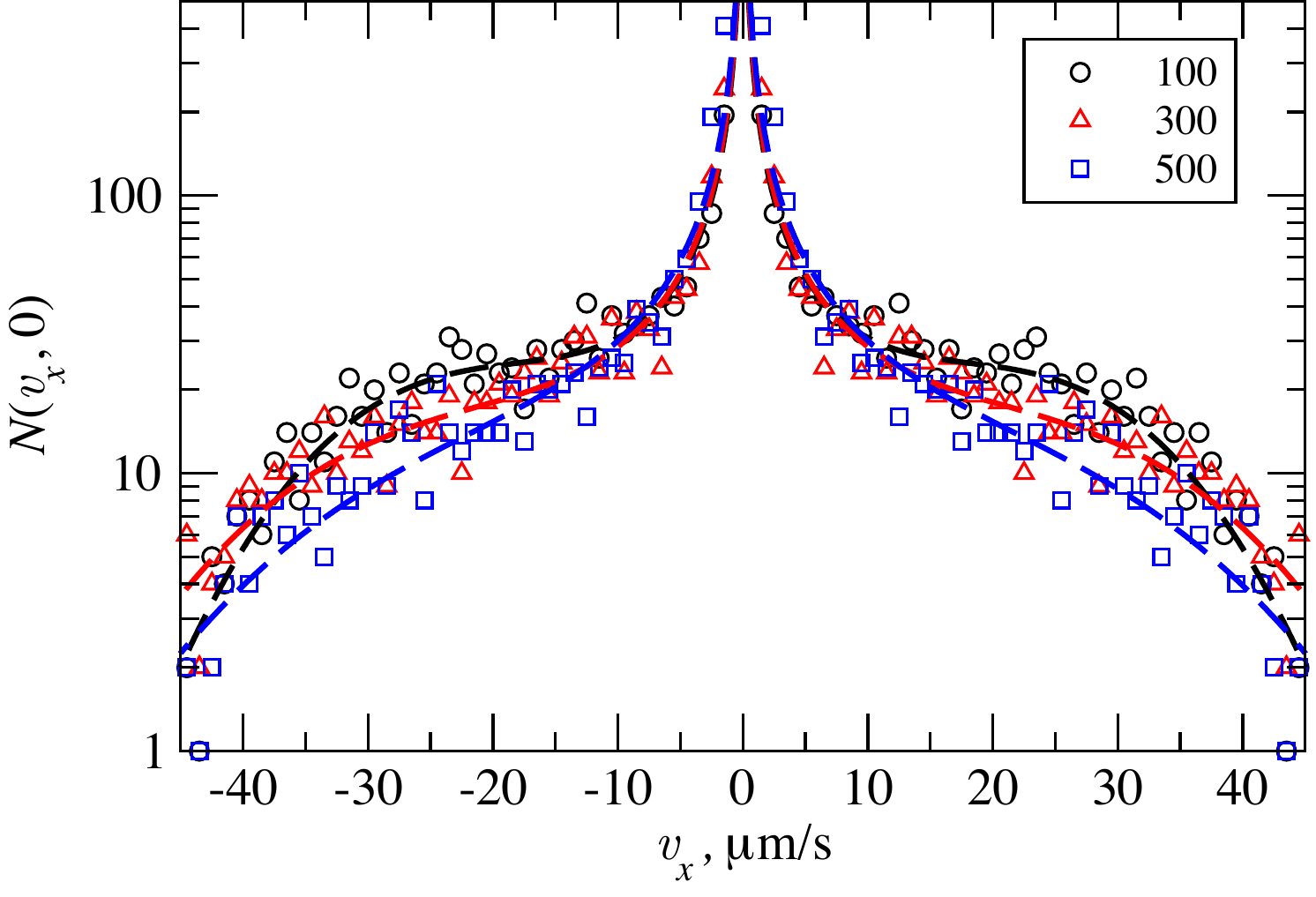}
\caption{\emph{Euglena gracilis} one-dimensional velocity histograms collected in intense blue light (points) and the fits with Eq. \eqref{eq:depot_cart} (lines). The labels on the experimental sets indicate the time interval within which the data were collected: 100 corresponds to the 0-100 seconds, 300 to  200-300 seconds, and 500 to 400-500 seconds from the beginning of the experiment.%
\label{fig:comparison}}
\end{figure}

We now compare the observed {\it Euglena} cell velocity histograms with the
predicted shape of the distribution. The histograms along with the
theory curves are presented in Fig. \ref{fig:comparison}.
All the sets could be fitted equally well. We do not show all of them as they practically overlap.
The correlation coefficient is in the range from 0.944 to 0.955 in different sets.
We see that the cell motion matches very well the predicted shape for systems with
active noise of high amplitude as shown in Fig. \ref{fig:depot}. It has a characteristic strong peak around zero
velocity. The fit is good in the whole velocity range.
Moreover, we observe that the distribution changes with time during the
experiment. The peak close to zero velocity becomes even stronger with time while the wings of the distribution, corresponding to
the normally moving cells, get somewhat weaker, which is consistent with the increase of the fluctuation amplitude
$D_v$ (see curves with $D_v=0.3$ and 1.0 in Fig. \ref{fig:depot}, right panel).

We attempted to interpret the latter observation using the depot model
with active noise. The parameters of the motion are as shown in Table
\ref{table}. As we see from the data, the ratio of energy influx to
the fluctuation amplitude $q/(2 D_v)$ remains nearly constant during
the experiment. The ratio $\gamma/(2D_v)$ is decreasing by about
50\%. Although we cannot resolve $D_v$ independently, we can assume
that the viscous friction coefficient $\gamma$ stays constant as it
reflects just the shape of the cells. We determined it in our previous
work for \emph{Euglena} cells $\gamma = 1.2 \times 10^{-7}$ N/(m/s)
\cite{romensky.m:2014}. Then, the decrease of the ratio
$\gamma/(2D_v)$ indicates the increase of the fluctuation amplitude by
about 50\%. At the same time, as the ratio $q/(2 D_v)$ does not vary
much (the variation is 5\% -- 7\%), corresponding to values of $q =
1.8 \times 10^{-16}$ W to $2.7 \times 10^{-16}$ W, we can infer the
increase of the incoming power by ca. 45\%. The ratio $c/d$ is
increasing most significantly during the experiment, the total growth
being more than 100\%. The propulsion velocity that is
  defined in the depot model \cite{schweitzer.f:1998,romanczuk.p:2012}
  by Eq. (\ref{eq:depot_v_0}) stays fairly constant at first, $ v_0
  \approx 26$ $\mu$m/s, but decreases to 18 $\mu$m/s towards the
  end. In the last interval the difference $q/\gamma - c/d $ turns
  negative and the propulsion velocity $v_0$ as parameter looses its
  meaning. The second term $c/d$ expresses the rate of internal energy
  dissipation to energy transfer. Hence, the observation is consistent
  with the increase of internal dissipation in the cells during the
  experiment. The overall picture can be interpreted as follows: The
\emph{Euglena} cells exposed to intense illumination show negative
phototaxis and are getting more and more fatigued at long exposures:
The amplitude of velocity fluctuations as well are internal
dissipation are growing while the fraction of energy that is converted
into persistent motion is decreasing.
\begin{table}
  \centering
  \begin{tabular}{ccccc}
    Time label & $q /(2 D_v) $& $10^{-9} \gamma/(2 D_v)$ & $10^9 c/d$ & $v_0$, $\mu$m/s \\
    \hline
    100 & 11.2 & 5.6 & 1.3 &   25.9 \\
    200 & 10.7 & 5.2 & 1.3 &   26.6 \\
    300 & 10.7 & 3.8 & 2.1 &   26.9 \\
    400 & 10.7 & 4.0 & 2.2 &   22.0 \\
    500 & 10.5 & 3.3 & 2.9 &   17.7 \\
    600 & 10.4 & 3.8 & 2.9 &   N/A  \\
    \hline
  \end{tabular}
  \caption{Parameters of the depot model with active fluctuations from fit of Eq. (\eqref{eq:depot_cart}) to \emph{Euglena gracilis} one-dimensional velocity histograms.  There is no detectable persistent motion of the cells for the given damping $\gamma$ at time 500 to 600 seconds as $q/\gamma - c/d < 0$.}
  \label{table}%
\end{table}

\subsection{\label{sec:mult} Active multiplicative noise}

Here, we also demonstrate the application of the active noise concept to a model
with multiplicative noise, which has been reported for random migration of the social
amoeba {\it Dictyostelium discoideum} \cite{bodeker2010}. The authors of this study measured the first
two moments of the velocity increments per unit time as function of
the velocity parallel and perpendicular to the direction of motion.
From those data they approximated the propulsion function and
the noise term of a corresponding Langevin equation.

In detail, they proposed based on their measurements, a simple linear friction term combined with a multiplicative noise term, with linear dependence on speed. Thus, the motion of the cells in their model is entirely driven by fluctuations without a deterministic component in the driving force. Please note that the authors considered in their model only passive noise, independent on the heading of the cell.

Here we go beyond the friction model suggested in  \cite{bodeker2010} and calculate the speed distributions of self-propelled agents with active multiplicative noise and so-called Schienbein-Gruler friction function \cite{schienbein.m:1993,romanczuk.p:2012}. The function includes a constant driving force along the heading direction
\begin{align}
\label{eq:SG}
\bs{a}_p(t)&= \gamma v_0 \bs{e}_h(t).
\end{align}
and a linear damping by a Stokes friction $-\gamma v$ (c.f. Eq. \ref{eq:headingdyn}). Thus for $v_0=0$ the model reduces to the situation observed in {\it Dictyostelium} migration \cite{bodeker2010}. However, for $v_0>0$, it also accounts for the case of individual agents with a finite preferred speed.
Fluctuations along the heading direction depend on the velocity, whereas
angular fluctuations are assumed to have constant intensity, i.e.
\begin{align}
\label{eq:noise2}
\bs{\eta}_\text{a}(t)&=\sqrt{2D_v} g(\bs{v})\xi_v(t)\bs{e}_h+ \sqrt{2 D_{\varphi}} \xi_{\varphi}(t)\bs{e}_{\varphi} \ .
\end{align}
Here, $\xi_i(t)$ with $i,j=v,\varphi$ are again sources of independent
Gaussian white noise defined in Eq.(\ref{eq:noise}).
In agreement with \cite{bodeker2010} we consider the case
\begin{align}
\label{eq:noise3}
g(\bs{v})=(1+\gamma_1 |v|).
\end{align}
and $v$ is the velocity component along the heading direction. Thus,
the multiplicative noise grows along the heading direction in both
directions. We would like to underline that the additive part in the
noise for $v=0$ together with the Gaussian character of this noise can
perform transitions from a motion with a positive velocity along the
heading direction to a motion with negative one and vice versa.

Using the Stratonovich interpretation \cite{anishchenko.v:2002} we
can now calculate the stationary speed distribution. The distribution
$p_a(v,\varphi,t|v_0,\varphi_0,t_0)$ of the projected velocities $v$
along the heading direction and on its angle $\varphi$ obeys the
Fokker-Planck equation
\begin{align}
  \frac{\partial p_a}{\partial t}&=
  -\frac{\partial}{\partial v}\gamma (v_0-v)p_a + D_v
  \frac{\partial}{\partial v}(1+\gamma_1 |{v}|)\frac{\partial}{\partial
    v}(1+\gamma_1 |{v}|)p_a+
  \frac{D_{\varphi}}{v^2}\frac{\partial^2}{\partial \varphi^2}p_a\,.
\end{align}
The asymptotic stationary distribution $\lim_{t_0 \to -\infty}
p_a(v,\varphi,t|v_0,\varphi_0,t_0) = p_a(v)$ becomes independent on
the initial conditions and the angle. With continuity of the velocity
distribution function density $p_a(v)$ at the origin $v = 0$ we obtain
\begin{align}
\label{eq:stat1}
p_a(v)& \propto (1+\gamma_1 |v|)^{(-1+\frac{\gamma}{D_v\gamma_1^ 2}+\frac{\gamma v_0}{D_v \gamma_1}\frac{v}{|v|})}\exp\left(-\frac{\gamma}{2 D_v\gamma_1}|v|\right).
\end{align}


\begin{figure}[t]
  \centering
  \includegraphics[width=\linewidth]{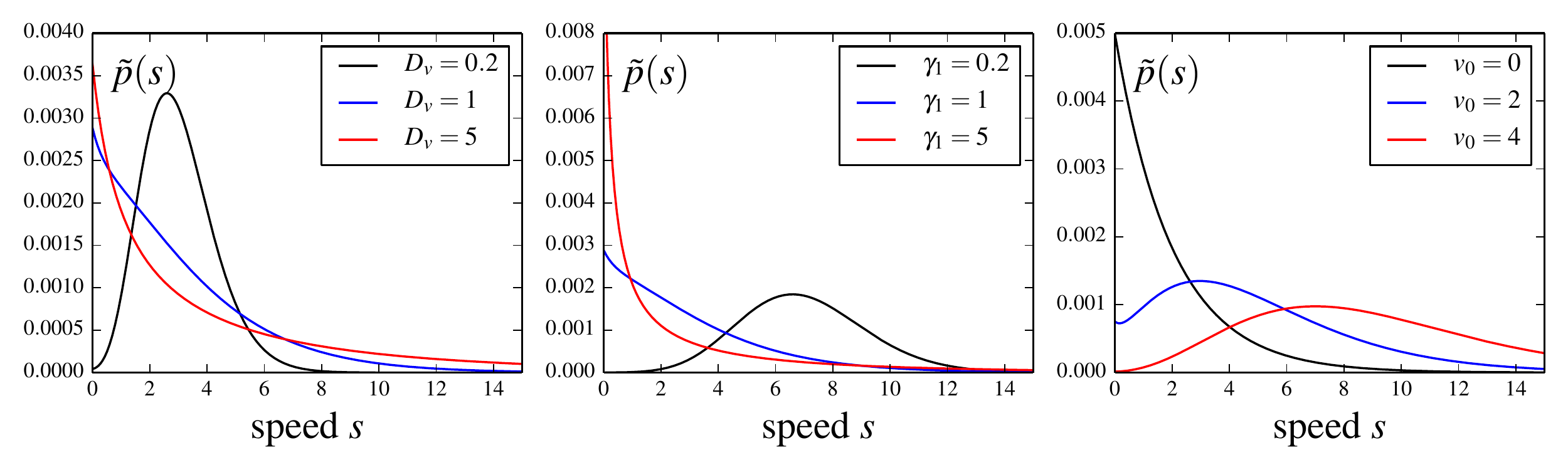}
  \caption{Stationary speed distribution $\tilde p(s)$ Eq. (\ref{eq:stat2}) for the
    model with multiplicative noise Eq. (\ref{eq:noise3}) for different values of $D_v$ (left), $\gamma_1$ (center) and $v_0$ (right). Default parameters if not varied:
    $\gamma=1.0$, $\gamma_1=1.0$, $D_v=1.0$, $v_0=1.0$.}
  \label{fig:mult1}
\end{figure}
\begin{figure}[t]
  \centering
  \includegraphics[width=\linewidth]{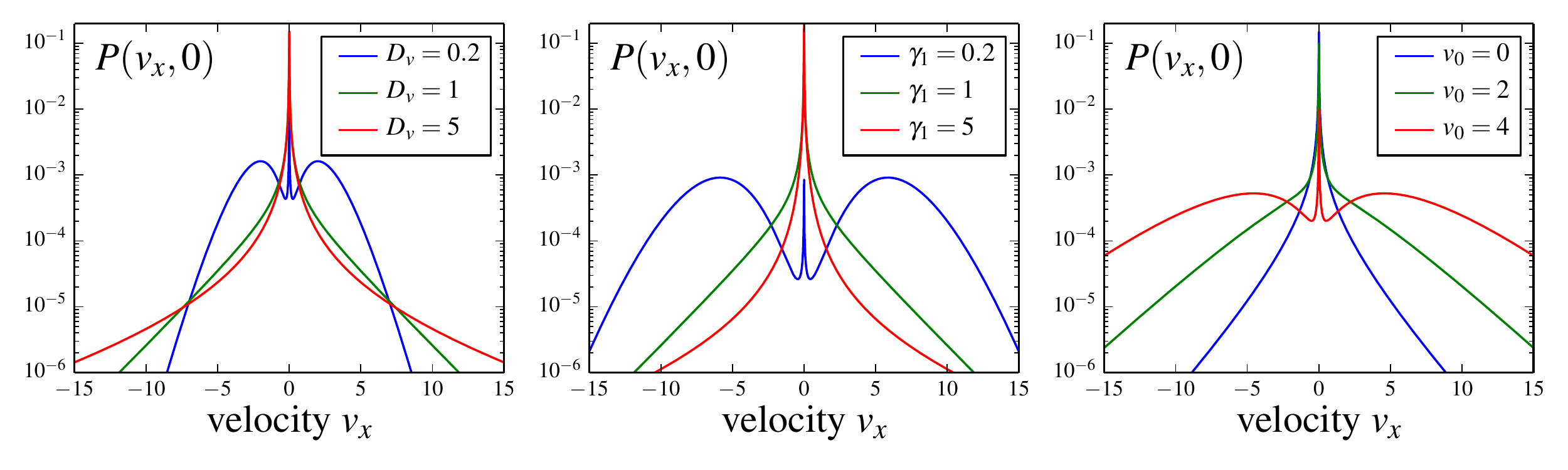}
  \caption{Stationary velocity distribution of the $v_x$-velocity component $P(v_x,0)$, Eq. (\ref{eq:stat3}) for the
    model with multiplicative noise Eq. (\ref{eq:noise3}) for different values of $D_v$ (left), $\gamma_1$ (center) and $v_0$ (right). Default parameters if not varied:
    $\gamma=1.0$, $\gamma_1=1.0$, $D_v=1.0$, $v_0=1.0$.}
  \label{fig:mult2}
\end{figure}

Due to the lack of symmetry of the propulsion $v_0>0$, the probability
density function in $p_a(v)$ is also asymmetric with respect to the
origin $v=0$. Since the noise depends on the absolute value $|v|$, it
is symmetric and grows in negative and as well positive direction.
This dependence induces different slopes for negative and positive
velocities around $v=0$ and $p_a$ is not differentiable at the origin
for $v_0>0$.

Summation of probability at positive and negative velocities $v$
values yields the speed distribution
\begin{align}
\label{eq:stat2}
\tilde{p}(s)\propto (1+\gamma_1 s)^{(-1+\frac{\gamma}{D_v\gamma_1^
    2})}\exp\left(-\frac{\gamma}{2 D_v\gamma_1}s\right)
\left((1+\gamma_1 s)^{(\frac{\gamma v_0}{D_v \gamma_1})}+(1+\gamma_1
  s)^{-(\frac{\gamma v_0}{D_v \gamma_1})}\right)
\end{align}
Interestingly, the asymptotic behavior for large speed values decays
exponentially and not as a Gaussian. In Fig. \ref{fig:mult1} we show
examples of the speed distribution given in Eq. (\ref{eq:stat2}) for
different parameter sets.

Finally, we transform the above results to Cartesian coordinates and
obtain the two-dimensional Cartesian velocity probability density,
which diverges at the origin:
\begin{align}
\label{eq:stat3}
P(v_x,v_y)=\frac{1}{\sqrt{v_x^2+v_y^2}}\tilde{p}(s=\sqrt{v_x^2+v_y^2})
\end{align}
The corresponding graphs of this function for $v_y=0$ are shown in
Fig. \ref{fig:mult2} for different values of $D_v$, $\gamma_1$ and
$v_0$ (as in Fig. \ref{fig:mult1}). We stress that as in the case of additive noise, we obtain a pronounced peak of
  the velocity distribution at vanishing velocities.

\section{\label{sec:conclusion} Conclusions}

In this work, we have discussed the concept of self-propelled
particles with active fluctuations in the context of experimentally
measured velocity distributions of \emph{Euglena gracilis} algae. In
biological agents, the locomotion is controlled by multiple internal
decision processes, which cannot be easily accessed. These complex,
unresolved influences will strongly contribute to the apparent
stochasticity of the motion, and in the first approximation they can
be conveniently summarized as active fluctuations. The intensity of
such active fluctuations and its dependence on various environmental
parameters yield important information on the internal states of the
biological agents and determines ecologically relevant quantities as
for example the spatial dispersal (diffusion) of individuals
\cite{romanczuk_active_2011}. Here, our results show that the depot
model \cite{schweitzer.f:1998} of active Brownian motion together with
active additive fluctuations perfectly describes experimentally
measured velocity distributions. The fitting of theoretical
distributions to experimental data reveals a systematic change of
locomotion parameters of individual algae with the duration of the
experiment, which provides important insights into the adaptation of
individual cells to a stressful environment.

We should add that the \emph{Euglena} single-cell algae might demonstrate a
more complicated behavior as it was reported previously. It might be
one of the simplest systems showing fatigue in response to external
stress.  Although we successfully described the population statistics,
on the single cell level the mechanism of the active fluctuations and
of the observed ``fatigue'' remains unclear. This mechanism can be
investigated further if one addresses complex and not fully understood
impulse-response behaviors involving the photoactivated adenylyl
cyclase \cite{Ntefidou2003,Ntefidou} and protein kinase \cite{Daiker}.

The theory of active fluctuations predicts a divergence of
two-dimensional Cartesian velocity probability densities irrespective
on the propulsion details \cite{romanczuk.p:2011}. The corresponding
diverging probability density manifests itself by a finite (local)
maximum at the origin, as it is the case in our experimental data.
Similar maxima in Cartesian distributions have been observed also in
very different experimental systems such as migrating {\it
  Dictyostelium discoideum} amoeba \cite{li_persistent_2008} or
self-propelled colloids \cite{ke_motion_2010}. Thus we expect that
active fluctuations are ubiquitous in living and non-living active
matter systems and therefore should be taken into account in
interpretation of corresponding experimental results and their
theoretical description.

\section*{Acknowledgements}
This work has been supported by the Deutsche Forschungsgemeinschaft
via IRTG 1740 (L.~S.-G.). P.~R. acknowledges support by the DAAD via
the P.R.I.M.E. fellowship.

\bibliographystyle{spphys}
\bibliography{Active_Fluctuations}

\end{document}